# Efficient Thermal Conductance in Organometallic Perovskite $CH_3NH_3PbI_3$ Films

Qi Chen,[1] Chunfeng Zhang,[1,2,a] Mengya Zhu,[1] Shenghua Liu,[1] Mark E. Siemens,[3] Shuai Gu,[4] Jia Zhu,[4] Jiancang Shen,[1] Xinglong Wu,[1] Chen Liao,[5] Jiayu Zhang,[5] Xiaoyong Wang,[1] and Min Xiao[1,2,6,b]

[1]National Laboratory of Solid State Microstructures, School of Physics, Nanjing University, Nanjing 210093, China
[2]Synergetic Innovation Center in Quantum Information and Quantum Physics, University of Science and Technology of China, Hefei, Anhui 230026, China
[3]Department of Physics and Astronomy, University of Denver, 2112 East Wesley Avenue, Denver, Colorado 80208, USA
[4]School of Engineering and Applied Science, Nanjing University, Nanjing 210093, China
[5]Advanced Photonic Center, Southeast University, Nanjing 210096, China
[6]Department of Physics, University of Arkansas, Fayetteville, Arkansas 72701, United States

E-mail: a) cfzhang@nju.edu.cn; b) mxiao@uark.edu

**Abstract:**

Perovskite-based optoelectronic devices have shown great promise for solar conversion and other optoelectronic applications, but their long-term performance instability is regarded as a major obstacle to their widespread deployment. Previous works have shown that the ultralow thermal conductivity and inefficient heat spreading might put an intrinsic limit on the lifetime of perovskite devices. Here, we report the observation of a remarkably efficient thermal conductance, with conductivity of 11.2±0.8 W m$^{-1}$ K$^{-1}$ at room temperature, in densely-packed perovskite $CH_3NH_3PbI_3$ films, via noncontact time-domain thermal reflectance measurements. The temperature-dependent experiments suggest the important roles of organic cations and structural phase transitions, which are further confirmed by temperature-dependent Raman spectra. The thermal conductivity at room temperature observed here is over one order of magnitude larger than that in the early report, suggesting that perovskite device performance will not be limited by thermal stability.

The past few years have witnessed a meteoric rise in research on solar cells with organometallic perovskite semiconductors[1-21]. Solar cells with power conversion efficiencies up to ~ 20%[22] have already been achieved in solution-processed perovskite devices benefiting from the long-range balanced carrier diffusion[23-25] and low density of midgap defects[25] in these perovskite compounds. These organometallic compounds also show promise for applications beyond photovoltaics, such as lasers[26-31], light-emitting diodes[32-35], photocatalyst[36], and gamma-ray sensitizers[25]. In spite of great successes in device demonstrations in labs, the real potential for practical outdoor applications of these devices is currently questioned by stability concerns[17,37].

Heat dissipation is a ubiquitous concern affecting the lifetime of any optoelectronic device, and it is especially important in the perovskite-based devices because of the potential material degradation at higher temperature[17]. However, while great efforts have been made to understand the electrical and optical properties of perovskite semiconductors, very few studies have targeted their thermal properties[38-40]. In the primary study to date, thermal conductivity in perovskite methylammonium lead iodide ($CH_3NH_3PbI_3$) samples was measured by a steady-state transport approach, and was reported to be very low ( 0.3 W $m^{-1}$ $K^{-1}$) at room temperature[39] – raising the concern that inefficient thermal conductance may prevent fast heat spreading and limit the lifetime of perovskite-based optoelectronic devices[39].

In this paper, we report the observation of a much larger thermal conductance in non-porous samples of polycrystalline $CH_3NH_3PbI_3$ films. Using the non-contact time-domain thermo-reflectance (TDTR) technique, we find the thermal conductivity of densely-packed $CH_3NH_3PbI_3$ films to be 11.2±0.8 W m$^{-1}$ K$^{-1}$ at room temperature. The measured thermal conductivity exhibits a glass-like temperature-dependent behavior below 150 K, implying an important role played by the disordered organic cations in the perovskite polycrystalline films. Moreover, the structural phase transitions of $CH_3NH_3PbI_3$ are manifested as anomalous temperature-dependent jumps in thermal conductivity at 140 - 160 K and ~ 330 K. The efficient thermal conductivity at room temperature observed here, which is over one order of magnitude larger than the value reported earlier, suggests that thermal stability is not a limiting factor in perovskite optoelectronic devices.

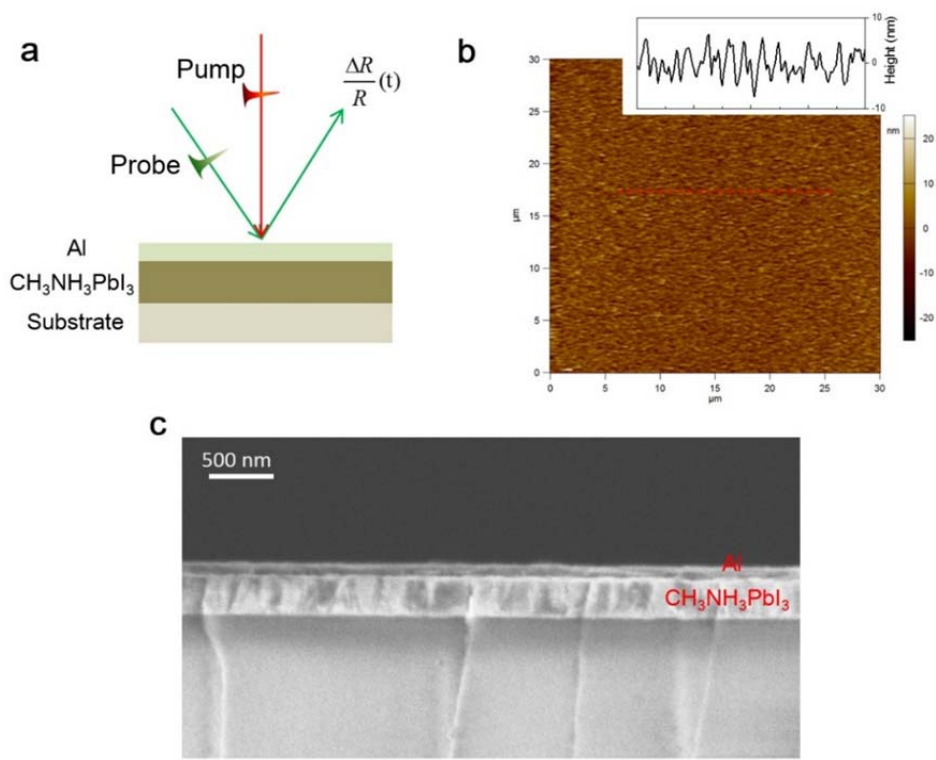

**Figure 1**. $CH_3NH_3PbI_3$ perovskite sample characterizations. (a) Schematic diagram of the TDTR measurement. (b) AFM image of the perovskite film in an area of 30×30 μm$^2$. The inset shows the height profile measured along the dashed line in the AFM image. (c) Cross-sectional image of the structure of Al-coated perovskite film used for TDTR measurements.

TDTR is well established as a reliable non-contact technique for characterizing the thermal conductivities of bulk and nanostructured materials[41-44]. As schematically shown in Fig. 1a, a metal film is deposited on top of the perovskite film sample as a thermal transducer. An ultrashort laser pulse heats the metallic layer in a few picoseconds, and the subsequent heat conducted into the perovskite diffuses inside the perovskite material, causing a cooling of the metal layer. This time-dependent behavior is monitored by the temperature-sensitive reflectivity of the metal surface using probe pulses that are delayed with respect to the pump pulses. To avoid heat accumulation during the measurement, we use a laser of low-repetition rate at 1 kHz as established in literature[45,46]. The power of the 800 nm pump laser is set at a low value (~ 0.2 mJ cm$^{-2}$) to ensure a linear response. To confirm the procedure in our measurement, a silicon wafer has been employed as a reference sample so that our measurements can be calibrated with established work[41,47].

The TDTR measurement relies on the reflectivity of the metal surface, so it is critical to have a flat film of perovskite material to minimize the light scattering. To meet this demand, we adopt the approach of solvent engineering modified from the procedure introduced by Jeon *et al.* to prepare the perovskite films[7]. With this improved approach, the surface roughness of a perovskite film can be optimized to be on a scale of less than ~ 10 nm as characterized by atomic force microscopy (AFM), ensuring a flat surface after metallic coating (Fig. 1b). The crystalline structure and electronic band structure of the samples have been confirmed by x-ray diffraction and photoluminescence spectroscopy (Supplementary Fig. 1 and 2). We characterize the cross section of the multilayer structure with scanning electron microscopy (SEM)

(Fig. 1c). SEM shows that the films are densely packed with no obvious porous structure observed and the interface between the perovskite layer and metal film is distinct with good uniformity. The resultant structure after aluminum deposition shows a reflective and smooth surface with excellent optical quality, enabling a reliable quantification of thermal conductivity of perovskite samples.

Figure 2(a) plots a typical TDTR curve of an aluminum-coated perovskite film recorded at room temperature. The early-stage fast decay reveals information about the electron-phonon coupling in the metal film and thermal conductance at the interface between the metallic layer and perovskite layer. Thermal conductance of the perovskite material governs the late-stage dynamics (> 200 ps). To extract the value of thermal conductivity, the thermal decay profile is analyzed with a one-dimensional thermal transport equation[47,48] in the form of $\partial T / \partial t = \alpha (\partial^2 T / \partial x^2)$ for a multilayer system using the well-established Clark-Nicoson finite difference scheme (see Supplementary Information for more details)[41,42,45], where $T$ is the temperature, $x$ is the distance normal to the surface, and $\alpha$ is the thermal diffusivity that can be employed to calculate the thermal conductivity. The validity of the procedure is confirmed by performing a control experiment on a silicon wafer; the measured thermal conductivity for silicon of 148±7 W m$^{-1}$ K$^{-1}$ is consistent with the literature value[41,47].

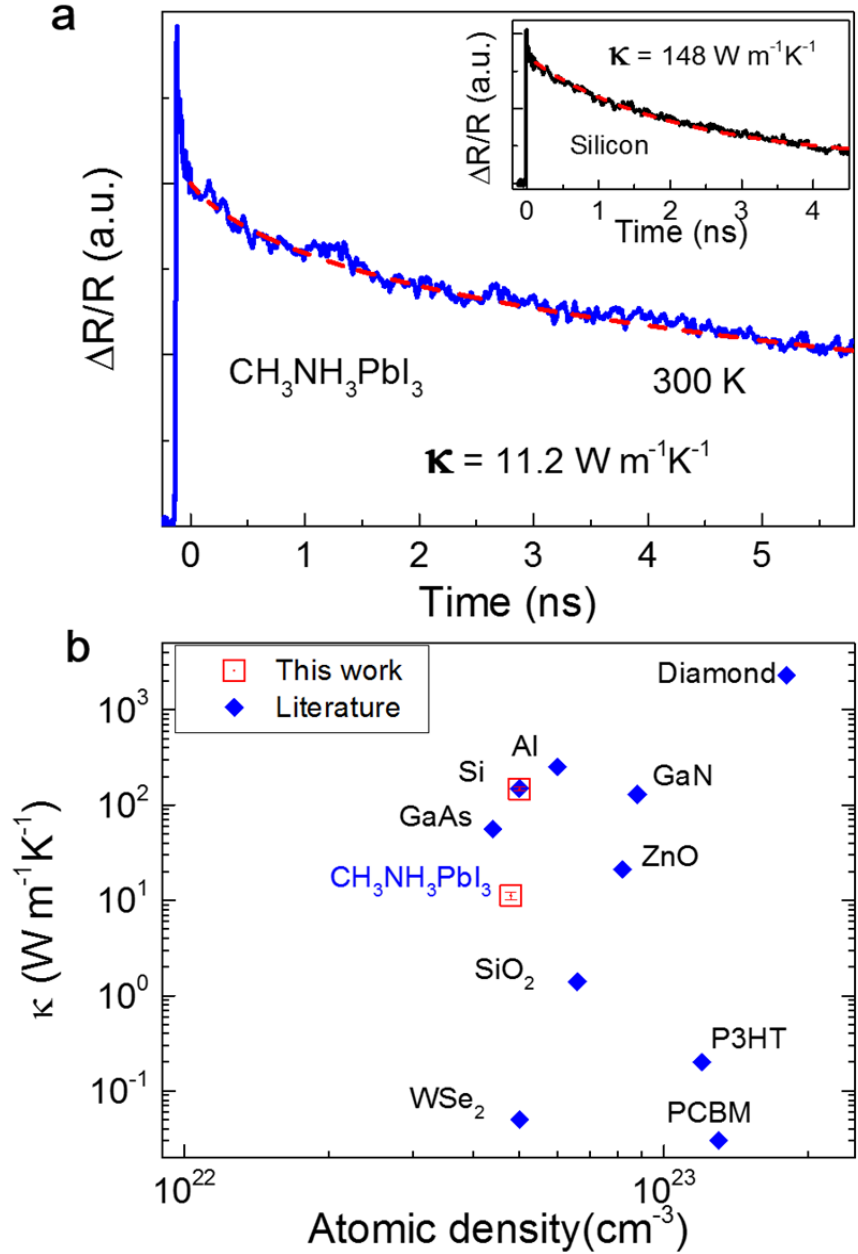

**Figure 2.** Thermal conductivity at room temperature. (a) A typical TDTR trace and the best-fit thermal model for the $CH_3NH_3PbI_3$ film recorded at room temperature. The inset shows the result of a reference experiment on silicon. (b) Room temperature thermal conductivities of various materials plotted as a function of their atomic density. The values for diamond, aluminum, silicon, $SiO_2$, GaAs, GaN, and ZnO are from Ref.[47,49]; P3HT and PCBM from Ref.[49]; and $WSe_2$ from Ref.[42].

For the experimental data recorded on the perovskite sample, the value of thermal conductivity is estimated to be 11.2±0.8 W m$^{-1}$ K$^{-1}$ at room temperature (Fig. 2(a)), which is much higher than the previously-reported result[39]. In Fig. 2(b), we plot this value of the $CH_3NH_3PbI_3$ film along with the reported thermal conductivities of various materials as a function of atomic density. Our measured thermal conductivity of the CH$_3$NH$_3$PbI$_3$ films is still low as compared with many materials of highly efficient thermal conductance, but the value is within an order of magnitude of those measured in some typical optoelectronic semiconductors like GaAs and ZnO (Fig. 2b), suggesting that thermal transport in the perovskite materials is actually efficient enough for standard device purposes. Nevertheless, the ultralow thermal conductivities in organic compounds of fullerene derivatives used in many perovskite devices may limit the thermal dissipation[49,50]. Other than that, the thermal conductance should not be a major issue affecting the stability of perovskite-based optoelectronic devices.

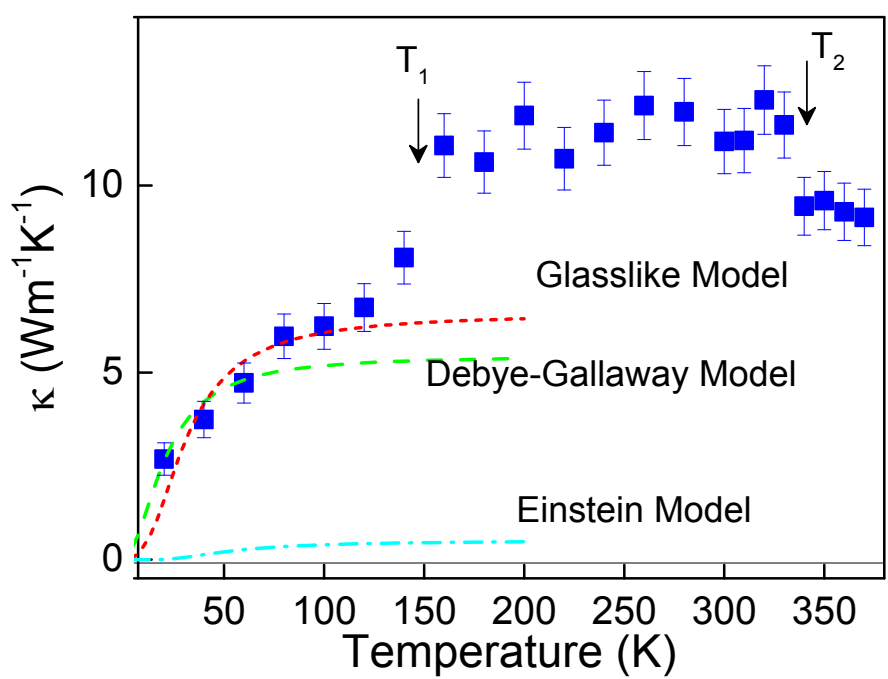

**Figure 3.** Temperature dependence of thermal conductivity. The measured thermal conductivity of the perovskite CH$_3$NH$_3$PbI$_3$ film is plotted as a function of temperature. The curves show fits to the low-temperature data with the Debye-Gallaway and glasslike models, and the calculated thermal conductivity with the Einstein model.

In semiconductors, the major heat carriers are long-wave acoustic phonons whose scattering affects the thermal transport[48]. The presence of disordered organic cations in the perovskite materials may cause divergence in physical mechanisms from the conventional semiconductors[51-53]. We have recorded the temperature-dependent thermal conductivity (Fig. 3) and compared the experimental results with three different models derived from material systems with different degrees of disorder (see Supplementary Information for more details). The Debye-Gallaway model that describes thermal transport in ordered crystalline systems[54] like silicon and germanium can only reproduce the data below 60 K (Fig. 3). The Einstein model[55], which considers a fully disordered material system, predicts the thermal conductivity to be more than one order of magnitude lower than the experimental data (Fig. 3). The temperature dependence of thermal conductivity up to 150 K is best described by the glass-like model[56] considering resonant scattering from two optical modes with frequencies of 70 cm$^{-1}$ and 230 cm$^{-1}$ (Fig. 3) that are comparable to the calculated frequencies of librational and torsional phonons relevant to the organic cations[52]. These results suggest that an intermediate degree of disorder relevant to organic cations best describes thermal transport in the $CH_3NH_3PbI_3$ films.

When increasing temperature beyond 150 K, significant departures from the thermal model appear (Fig.3), which are likely caused by structural phase transitions. It is known that the perovskite material $CH_3NH_3PbI_3$ undergoes two structure transitions above 150 K, one from orthorhombic to tetragonal phase at $T_1 \sim 162$ K and another from tetragonal to cubic phase at $T_2 \sim 330$ K[57]. Such phase transitions significantly modify the dispersion of acoustic phonons as well as the frequencies of librational and torsional modes of organic cations[52], suggesting the possibility of abrupt changes in thermal conductivity at those temperatures.

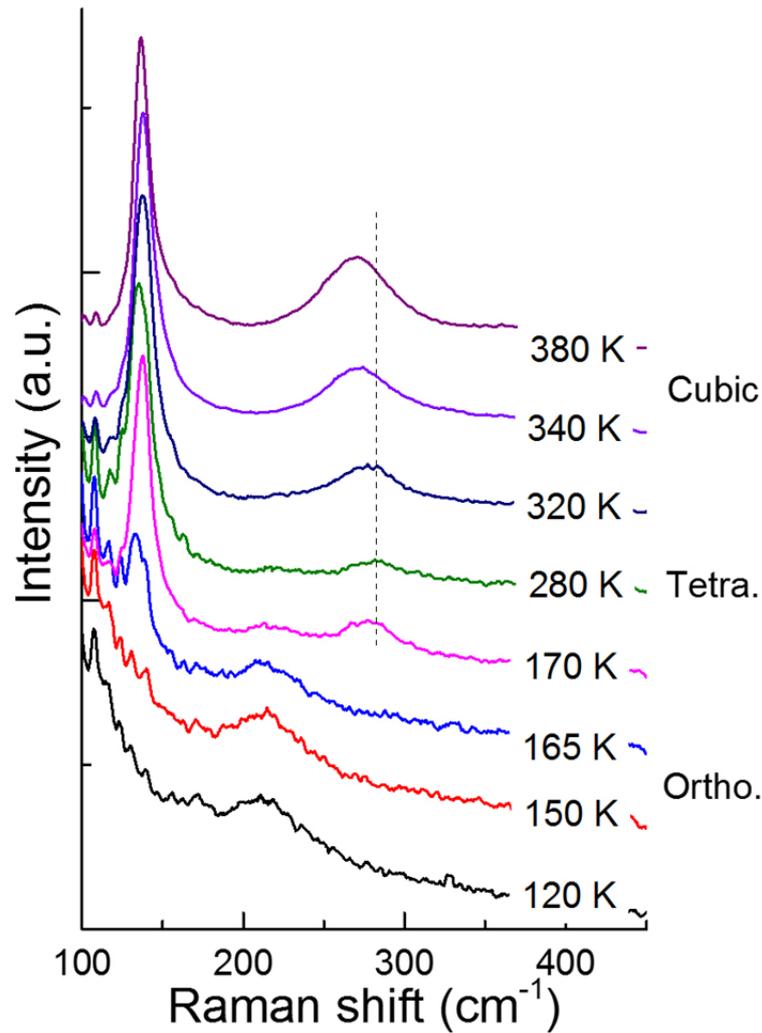

**Figure 4.** Raman spectroscopy of the perovskite film. The Raman scattering spectra of the CH$_3$NH$_3$PbI$_3$ film are recorded at different temperatures, indicating the phase transitions from orthorhombic to tetragonal phase at ~ 165 K and from tetragonal to cubic phase at ~ 330 K, respectively.

For confirmation, we have recorded the Raman scattering spectra as a function of temperature (Fig. 4), which show clear evidences of phase transitions at these temperatures. A broad peak at 215 cm$^{-1}$ with a linewidth of ~ 40 cm$^{-1}$ is distinct below 160 K, which diminishes when increasing temperature over 165 K. In the same temperature range, a broad peak at 280 cm$^{-1}$ and a sharp peak at 137 cm$^{-1}$ emerge. These changes are signatures of the transition from orthorhombic to tetragonal phase[52].

The peaks at 215 cm$^{-1}$ and 280 cm$^{-1}$ with broad linewidths can be assigned to the torsional mode[52], and the sharp peak at 137 cm$^{-1}$ is likely to be the librational mode of the tetragonal phase[52]. A frequency change of the torsional mode is also observed at ~ 330 K due to the transition from tetragonal to cubic phase. Besides this clear evidence from Raman spectroscopy, the phase transitions are also manifested in temperature-dependent photoluminescence spectra (Supplementary Fig. 2)[58,59]. These results can qualitatively explain the observed abnormal changes in the temperature dependence of thermal conductance.

The thermal conductivity value measured here is ~ 40 times larger than that measured in a polycrystalline sample of the same material by using the steady-state technique[39]. This distinct divergence could result from significant differences between the perovskite samples. The previous measurements were on porous structures inside polycrystalline samples[39] made from mechanically-pressed small crystals, which may induce thermal insulation; this could cause a significant underestimation of the thermal conductivity from well-prepared perovskite materials. In fact, our own measurements on porous samples prepared by the conventional approach show a much less efficient thermal conductance (Supplementary Fig. 5). Moreover, from the fitting procedure, the parameter (Supplementary Table 1) describing the coupling to the optical modes is one order of magnitude smaller than that estimated in the previous work[39], suggesting that the disorder of organic cations plays a reduced role (comparable to the boundary scattering) in thermal diffusion at low temperature in our sample. Such a speculation seems to be reasonable since the degree of structure distortion for organic cations is strictly limited for an orthorhombic structure in the perovskite sample at low temperature[51].

Based on the above discussions, we conclude that the thermal conductance in perovskite is dominated by the disorder of organic cations, which is probably a major factor that restricts the thermal transport in the hybrid perovskite materials to be less efficient than some inorganic optoelectric semiconductors. The scattering of grain boundaries in the polycrystalline materials could also constrain the thermal conductivity[60]. In light of these factors we can conclude that, its thermal conductance is efficient enough to achieve stability in optoelectronic devices with available technologies of thermal management. Considering the similar structures, it is natural to expect similar thermal behaviors in other members of the perovskite family of materials. Once the chemical stability is achieved, efficient perovskite devices with robust stability should be realizable, as evidenced preliminarily by a recent demonstration of fullerene-free perovskite solar cells with a lifetime of more than $10^3$ hours[8].

ACKNOWLEDGEMENTS

This work is supported by the National Basic Research Program of China (2013CB932903 and 2012CB921801, MOST), the National Science Foundation of China (91233103, 11574140, 11227406 and 11321063), and the Priority Academic Program Development of Jiangsu Higher Education Institutions (PAPD). We acknowledge Dr Xuewei Wu for his technical assistant.